\documentstyle [12pt] {article}
\begin{document}
\title { Geometrization of   Econophysics : An Alternative Approach for Measuring  Elements of Risk Management of an Economic System }
\maketitle
\begin{center}
{\bf {M.E.Kahil{\footnote{October University for Modern Sciences and Arts, Giza, EGYPT\\ Egyptian Relativity group (ERG) \\ The American University in Cairo, New Cairo, Egypt \\
e.mail: kahil@aucegypt.edu}}} }
\end{center}

\begin{abstract}The relationship between micro-structure and macro-structure of complex systems using information geometry  has been dealt by several authors. From this perspective, we are going to apply it  as a geometrical structure connecting both microeconomics and macroeconomics .  The results  lead us to introduce  new modified quantities into  both micro-macro economics that enable us to  describe the link between them. The importance of such a scheme is to find out -with some accuracy- a new method  can be introduced for examining  the stability of an economic system. This type of requirement is expressed  by examining  the stability of  the equations of path deviations for some economic systems as described in a statistical manifold.
Such a geometization scheme of economic systems is an important step toward identifying risk management factors and so contributes to the growing literature of econophysics.
\end{abstract}

\section{Introduction}
It is well known that the problem of  risk management is related to socio-economic systems as well as to epidemiology. A central problem of risk management is the development of   forecasting  models  to regulate or even prevent future  incidents that might cause instability  throughout a whole system. The necessity to manage risk has led many to seek a deterministic model that will allow us to describe exactly -- within the limits of defined parameters -- such a predictive model.
The demanding for such a model may  begin with a system of ordinary  differential equations to examine the  evolution of the system.  Such  a system , looks so naive to handle some current problems of economic systems even if there is some capability to express it in terms of non-linear partial differential equations . One of these  is to search for such a way of thinking enables us to express any economical issue or financial situation by importing some ideas based on physics as it can be useful to revisit these problems by means of a new paradigm shift.

Our approach is  based on describing economic problems using information geometry. This type geometry  replaces every point in its manifold  by a probability distribution of  several correlated and mutually interacting events  of each other .  This application of information geometry allows us to examine in depth the correlation between the micro-economic elements as well as to describe  how these correlations evolve within the macro-economic space, thus providing a mechanism to connect micro and macro economic factors together. However, in the meantime such a goal is far fetched but this approach of thinking may express that for every micro-economic element is responsible for establishing its corresponding macro-economic one.  Thus,  it may drag our attention to speculate whether or not the microeconomic element may exactly describe the well known corresponding macroeconomic element.
In other words, we propose that each macroscopic element is due to several identifiable micro-economic elements and that this relationship can be given precise definition using information geometry. Details of this hypothesis will be further developed in forthcoming studies.

\section{Mathematical Modeling of  Economic Systems}

\subsection{  Deterministic Models  in  Using  Differential Equations } It is well known that one of ways of describing a  deterministic model is by means of introducing differential equations for the system to examine and predict its evolution at different times.
An  attempt  to the describe  such a primitive  model  of macroeconomic growth is as follows [1]
 \begin{equation}
 \frac{D K}{Dt} =  I(t) ,
\end{equation}
Such that
$$Y = C(t) + S(t)$$,  $$S(t) =I(t)$$ and $$S(t) = \mu Y(t)$$

$$ K(\tau) =\nu Y(\tau) $$

 where $Y(\tau)$ is the national income, $S(t)$ is
 the amount of compensation and accumulation per year, $I(t)$ is a  amount of investment per year; $K(t)$ is the amount of  capital  per year  and $\nu$ is an arbitrary  constant.  For a better mathematical formulation,  The above  model can be modified to follow Lotika-Volterra  is becoming  [2]:

 \begin{equation}
\frac{dK}{dt} =-\alpha_{1} KI + \alpha_{2} K
 \end{equation}
and
 \begin{equation}
\frac{dI}{dt} =\alpha_{1} KI + \alpha_{2} I
\end{equation}
where  $\alpha_{1} $  and  $\alpha_{2}$  are constant coefficients.
 However, these constants do not match with the current situation especially for dealing with a large set of data which may impose some stochastic parameters. This leads us to replace them by a family of smooth probability distribution functions having its own means and variance [3]
\subsection{Geometrization of  Economic Systems }

The concept of geometrization of macroeconomics stems from expressing every element which describes the case of macro-economy as a dimension: the more dimensions described in the space of macroeconomics, the greater precession of forecasting the behavior macro-economy. From this perspective, it will be important to start by defining the manifold of the macro-economy as a 2-dimensional curved space. The term curved is admitted to include the chaos in its contents.
Equations of geodesic and geodesic deviation of this space will enable us to examine the evolution of such a system and its tendency of stability through its corresponding deviation vector.

 In this work, we  geometrize  macroeconomics by proposing that all acting variables in an economic system, can be expressed in terms of  dimensions in a  manifold - economical manifold.  This technique may  be used analogously to describe the evolution of epidemics using  allometric spaces  [2]
 \begin{equation}
L=g_{ab}U^{a}U^{b}
 \end{equation}
where $U$ is the tangent vector with respect to the parameter $t$. However, in this  approach,  we are going to obtain path and path deviation equations from one single Lagrangian using the Bazanski Lagrangian
:
 \begin{equation}
L = g_{_{ab}}{U^{a}} {\frac{D \Psi^{b}}{Dt}}
 \end{equation}
where $a,b=1,2,3,..n$ and $\frac{D \Psi^{\alpha}}{Dt}$ is the covariant derivative with respect to a parameter $t$ and  the line element as defined  by Rao  [4]

 \begin{equation}  dS^2 = \frac{1}{\sigma_{i}^2} d \mu_{i}^2  +  \frac{1}{\sigma_{i}^2}d \sigma_{i}^2   \end{equation}
where, $\mu_{i}$ is the mean of the elements macro-economic space and  $\sigma$ is the standard deviation of the same elements.  Taking into consideration in that macroeconomic space $ K={K_{1}, K_{2},......K_(n)}  , I ={I_{1}, I_{2},.......I_{n}}$
in which the space enclosed   $(K,I) $is  the 2-dimensional macro-space of Capital -Income.

Taking the variation with respect to the deviation vector $\Psi^{c}$ and the tangent vector $U^{c}$ respectively one obtains
\underline{ the path equation}
 \begin{equation}
\frac{dU^{c}}{dt}+ \Gamma^{c}_{ab} {U ^{a}}{U^{b}}=0
 \end{equation}
and \underline {its path deviation equation}
 \begin{equation}
\frac{D^2 \Psi^{c}}{Dt^2}= R^{c}_{abd}U^{a}U^{b}\Psi^{d}.
\end{equation}

Thus, the path and path deviation equations of the  KI-model can be obtained from the following Bazanski Lagrangian  [5]
   \begin{equation}  L= g_{\mu \nu} U_{KI}^{\mu} \frac{D \Psi_{KI}^{\nu}}{Dt}
 \end{equation}
where $ U^{\mu}_{KI} = (K,I) $ and $ \Psi_{KI}^{\nu} = ({\Psi_{K}},{\Psi_{I} }) $.  Taking the variation with respect to the deviation vector $\Psi^ {\sigma} $we  get the following components of the path equation
\begin{equation}
\frac{d K}{ dt}+ \Gamma^{1}_{11}{ K}^2 + \Gamma^{1}_{22}{ I}^2 + 2\Gamma^{1}_{12} {K}{ I} =0,
 \end{equation}
and
 \begin{equation}
\frac{d I}{ dt}+ \Gamma^{2}_{11}K^2 + \Gamma^{2}_{22}{ I}^2 + 2\Gamma^{2}_{12} { K}{ I} =0.
 \end{equation}
And taking the variation with respect to velocity vector $U^{\sigma}$ we get
 the corresponding components of the  path deviation equation  [6] :
 \begin{equation}
\frac{D^2\Psi_{K}}{ Dt^2} =   R^{1}_{112} { K}^{2} \Psi_{I}  + R^{1}_{121}{ K} {I} \Psi_{K}  +R^{1}_{212} { K} { I} \Psi_{I}+ R^{1}_{221} I^{2} \Psi_{K}, \end{equation}
and
\begin{equation}
\frac{D^2\Psi_{I}}{ Dt^2} =   R^{2}_{112} { K}^{2} \Psi_{I}  + R^{2}_{121}{ K} { I} \Psi_{K}  +R^{2}_{212} { K} { I} \Psi_{I}+ R^{2}_{221} {I}^{2} \Psi_{K}. \end{equation}

Although this step of geometrization  elements of macroeconomics  is quite useful, it  is still unacceptable for describing the real factors of risk management issue. This due to lack of contact between the variables of macro economics and microeconomics. Such a unification can not be achieved without getting some geometrical techniques to express and relate both these domains. One possible approach is to use the geometrization method of econophysics by  applying information geometry of  maximum entropy method.
\section{ Econophysics: Economics  as Complex Systems  }
The interaction between physics and economics leads to study some irregular problems  such as high frequency finance, financial risk and some complex systems  using an interdisciplinary science called econophysics.

Econophysics was started in the mid 1990's .[7]  mainly dealing with   complex problems in  economics as well as  financial markets, to obtain relevant explanations for vague problems in both economics and finance e.g. on heterogeneous agents and far-from-equilibrium situations.     Accordingly, the notation of a complex system  may lead to express it by  means of thermodynamics-  They are systems of many interacting agents of highly nonlinear features. This amount of data allows a detailed statistical description of several aspects of the dynamics of asset price in a financial market. These results are based on some data of several complexity in the price of dynamics of financial assets.

  The thermodynamic model induces temperature and entropy. With no information about these variables, it is not possible to find the correct equilibrium conditions for the two systems.
The fundamental law of equilibrium statistical mechanics is Botlzamann-Gibbs  law, which states that the probability distribution of energy

 $E$ is $p(E)=C  e^{-E/T}$, where $C$ is a normalizing  constant and $T$ is an effective  temperature . The main ingredient that is essential for the derivation of Botlzmann-Gibbs law is the conservation of energy. Thus one may generalize that any conserved quantity in a big statistical system should have an exponential probability distribution in equilibrium [7].  It has been found that laws of money is responding to the same laws of energy. i.e., money can also be conserved   e.g.${ \it{financial potential energy + financial ki4netic energy = constant}} $   [8]  and its distribution is following the Boltzmann-Gibbs law :
 \begin{equation}  p(m) =C  e^{-m/ \ bar{T}}  \end{equation}
$$ \int^{\infty}_{0} p(m)dm =1$$ and,$$ \int^{\infty}_{0}m p(m)dm =M/N , $$  taking $C= \frac{1}{\bar T}$ and $\bar T = M/N$. due to $M=  n_{b}m^{b}$ . Here $m $ is money and $\bar T$ is  the average amount of money per economic agent which is analogous to the temperature in physical systems.

Let an economic system consists of $N$ agents, ........ the total income I(t) it corresponds to the sum of modes of income distribution between these agents as the statistical weight of the state with this income, using a characteristic function $n (I(t), N)$ for this task.

Now, it is possible to introduce the concept of equilibrium.We may consider  two systems are in equilibrium , if the function of income distribution remains constant, there is no flow of income  among agents appears.  Let one system with the total income $I_{1}$ have $N_{1}$ , the number of agents while  the other is $I_{2}$ with its number of agents $N_{2}$
If  the system  is composed by two subsystems $n_{1}(E_{1}, N_{1})$ and $n_{2}(E_{2}, N_{2})$, then the total income and number of agents become $E_{1}+E_{2}$. and $N_{1}+ N_{2}$ respectively. This may give rise to consider  that the state of equilibrium .

Let $\Delta I$ be a certain part of income which passes from sub- system(1) to sub- system (2), which produces a change in each statistical weight to become from $n_{1}(I_{1}-\Delta I, N_{1})$ to $n_{2}(I_{2 +\Delta E, N_{2}})$ .

According to the principle of equal probability, the most probable state   of these  subsystems  is the one  the greatest statistical one i.e.  the maximum of the function $n_{total}(E_{1}, E_{2}, N_{1}, N_{2})$.  This whole  system is regarded as its based on that the total income  remains  $E_{1}+E_{2}$ , without  transferring of agents, thus, the overall  statistical weight of  this system becomes   $$ n_{total}(E_{1}, E_{2}, N_{1}, N_{2}) = n-{1}(E_{1}, N_{1})  X n-{2}(E_{2}, N_{2})  $$
Considering in this case, $ E_{1}+ E_{2} =  constant $, then $$ \Delta E_{1}= -\Delta E_{2} .$$
From this perspective,  the maximum of  $n_{total}$ is the maximum of $ln (n_{total})$.Since,
$\ln n_{total} = \ln n_{1}+ \ln  n_{2}$
which gives the condition of maximum statistical weight is obtained  from the equilibrium condition :
   \begin{equation}  \frac{d}{d E_{1}}\ln n_{1}(E_{1}, N_{1}) = - \frac{d}{d E_{2}}ln n_{2}( E-E_{1}, N_{2})  , \end{equation}
$$ \Delta E_{1}= - \Delta  E_{2} $$
to give
$$ \frac{d}{d E_{1}}\ln n_{1}(E_{1}, N_{1}) =  \frac{d}{d E_{2}}\ln n_{2}(E_{2}, N_{2}).$$
If the two two systems have such a condition
 $$ \frac{d}{d E_{1}}\ln n_{1}(E_{1}, N_{1}) =  \frac{d}{d E_{2}}\ln n_{2}(E_{2}, N_{2}) =  \frac{d}{d E}ln n (E, N).$$
then one can regard it from a thermodynamical perspective the inverse of the temperature
i.e.
  \begin{equation}   \frac{d}{d E}\ln n (E, N)= \frac{1}{T} , \end{equation} and  the logarithms of the statistical weight  is called the entropy of the system.
Thus, in  a state of equilibrium the interacting systems should have the same temperature.
In order to connect the above phenomena with economics, it is worth mentioning that an economical system is in state of equilibrium if it is almost homogeneous and it does not imply flows from one subsystem to another. However,  the homogeneity  exists only if there is no separation into such parts so that no major income flows are noted. The system is the state of equilibrium when the two subsystems have the same temperature which can not be calculated without knowing the entropy of the system. [11]

\subsection{ Entropic Dynamics : Information Geometry }

 The way to recognize  details about the transition from one state to another for a given system is  entirely by examining the change in their probability distributions. The most  reliable information about the transition state is  reaching to its maximum entropy. The maximum entropy may be interpreted geometrically by the possible trajectory in a statistical manifold that describes its evolution. It can be regarded that the method of maximum entropy can transform the manifold of states into a metric space [11]. This means that the change between two different states can be expressed in terms of a distance between them  and this distance can be defined in a statistical manifold.

The underlying geometry of this space stems from considering that at each point of the space, there exists an  n-dimensional manifold, a micro-space [11] .

Let the micro-states of any economic  system be labeled by x, and let $q(x)dx$ be the number of micro-states in the range  $d(x)$ . Also, there exists  a macro-state defined by  $\Theta^{\alpha}$   stands for    the expected values  for $n_{\Theta}$ variables expressing the micro-state  in the following way:

  $$< a^{\alpha} > = \int  dx p(x) a^{\alpha}(x)= \Theta^{\alpha} $$

  where  variables $a^{\alpha}(x)$ $(\alpha= 1,2,…n_{A}),$.
 At each values of $\Theta^{\alpha}$  there is a set of coordinates, expressing the macro-space. such that
the set $\Theta$ defines the $2l$-dimensional space of macro-space of the states of the system,  the statistical manifold $M_{s}$.Thus, the probability distribution $p(x|\Theta)$ represents the prior information contained in $q(x)$  innovated by $\Theta^{\alpha}$  [ 12]
which can be obtained by maximizing the entropy
 \begin{equation}
s(p) =- \int dx p(x) \log \frac{p(x)}{q(x)}
 \end{equation}
The difference between two states $A^{\alpha}$ and $ A^{\alpha}+ d A^{\alpha}$ is given by a small value $d S^2$  defined in the following way
 \begin{equation}
dS^2 = \int dx p(x|\Theta) \frac{\partial \log {p(x|\Theta)}}{\partial \Theta^{\alpha}} \frac{\partial \log {p(x|\Theta)}}{\partial \Theta^{\beta}}
 \end{equation}
 A measure of distinguish-ability among macro-states of the statistical manifold is defined by assigning a conditional probability is belonging the statistical manifold to each macro-state. This kind of assignment endows the statistical manifold with a metric structure. Specifically, the Fisher-Rao information $g_{\\mu \nu} ( \Theta )$ i.e.
 \begin{equation} g_{\mu \nu} ( \Theta) = \int dx_{p}(x|\Theta) \partial_{\mu} log _{p}(x|\Theta) \partial_{\nu} log _{p}(x|\Theta) ,   \end{equation}
where $\mu, \nu =1,2,3,….2l $ and $ \partial_{\mu} = \frac{\partial}{\partial \Theta^{\mu}}$ defines a measure of distinguish-ability among macro-states on. the statistical manifold of $M_{s}$ .
It is well known to apply information geometry it  requires a metric $g_{ab}$ is symmetric and positive definite, and Fisher metric admits the following properties :\\ 1. Invariance under transformations of micro-variables [13].
 \begin{equation} p(x| \Theta) \rightarrow \hat{p}( \hat{x} | \Theta)= [  \frac{1}{\frac{\partial f}{\partial x}} p (x| \Theta )] . \end{equation}
2.Covariance under reparametrization of statistical macro-space,
 \begin{equation}{g_{ab}}  \rightarrow \hat{g_{ab}}= [\frac{ \partial \Theta^{c}}{\partial \hat{\Theta^{a}}}\frac{\partial \Theta^{d}}{\partial \hat{\Theta^{b}}}g_{cd} ( \Theta)] ,  \end{equation}
such that
 $$ \hat{g_{ab}( \hat{\Theta})} =\int dx \bar{p} ( x |  \hat{\Theta}) \bar{\partial}_{a}  \ln  \bar{p} ( x |  \hat{\Theta}) \bar{\partial}_{b}  \ln  \bar{p} ( x |  \hat{\Theta}) . $$
\section{Geomerization in Macroeconomics using Information Geometry}
This  geometrization scheme for describing macroeconomic growth models  uses the richness of information geometry- defining each point in the macro-state space as a world of micro-structure with both  the correlated and uncorrelated variables.
In each version geodesic and geodesic equations-working for examining the stability of the system will be different-as will be studied in future work.
In our  present work   we   focus   primarily  on some primitive models-toy ones- to rewriting the economical issues as a mere set of non ordinary differential  equations,that might be adaptable to examine the degree of chaos  by means of measuring the evolution of entropy in any economic system. It is  evident that we can not determine the evolution of micro-states  due to  insufficient data . Instead we can study the distance between total probability distributions with parameters $(\mu_{1}, \mu_{2}, \sigma_{2})$ and $\mu_{1} + d \mu_{1},\mu_{2} + d \mu_{2}, \sigma_{2} + d \sigma_{2}$- assuming that $\sigma_{1} =0$. Once the states of the system can be defined , then the problem of quantifying   the  difference between macro-states$ \Theta$  and $\Theta+d \Theta$ is described by a dimensionless distance between the two states $p ( \vec{x}| \vec{\Theta} ) $ and $ p(\vec{x}| \vec{ \Theta} + d \vec{\Theta})$:
 \begin{equation} dS^2 = g_{ij} d \Theta^{i} d \Theta^{j} , \end{equation}
where
$$ g_{ij} = \int d \vec{x} p( \vec{x}| \vec{\Theta}) \frac{\partial log p( \vec{x}| \vec{\Theta})}{\partial \Theta^{j}})\frac{\partial log p( \vec{x}| \vec{\Theta})}{\partial \Theta^{j}})$$
is the Fisher- Rao metric
$$
g_{ab} ( \mu_{x}, \mu_{y} ; r )= \frac{1}{\sigma^2} \left (  \begin{array}{ccc}  -\frac{1}{r^{2}-1}  & \frac{r}{r^{2}-1}  &  0   \\    \frac{r}{r^{2}-1} & -\frac{1}{r^{2}-1} & 0  \\  0 & 0 &    4  \end{array} \right ) $$
$$
g_{ij} =  \left (  \begin{array}{ccc}   \frac{1}{\mu^{2}_{1}}  & 0  &  0   \\    0 & \frac{1}{\mu^{2}_{2}} & 0  \\  0 & 0 &    \frac{1}{\sigma^{2}_{2}} \end{array} \right )
$$
to get its line element [13]
 \begin{equation}
  dS^{2}_{M^3D} = \frac{1}{\sigma^{2}}(d \mu^{2}_{x} + d \mu^{2}_{y}+ 4 d \sigma^{2}_{x} )  \end{equation}
and its  the non-vanishing affine connection becomes
$$ \Gamma^{1}_{13} = -\frac{1}{\sigma} , \Gamma^{2}_{23}= -\frac{1}{\sigma} = \Gamma^{1}_{32} ,$$ $$\Gamma^{3}_{11} =  \frac{1}{{4 \sigma}(r^{2} -1)} , \Gamma^{3}_{12}=\frac{r}{{4 \sigma}(r^{2} -1)} =\Gamma^{3}_{21},  \Gamma^{3}_{22}=  \frac{1}{{4 \sigma}(r^{2} -1)} , \Gamma^{3}_{33} = - \frac{1}{ \sigma} . $$
The geodesic equation  describes a reversible dynamics whose solution is the trajectory between initial $\Theta_{i}$l and final macrostate $\Theta_{f}$  which can be expressed in the following way
 \begin{equation} \frac{d^{2} \mu_{x}}{d S^2}- -\frac{2}{\sigma} \frac{\mu_{x}}{dS} \frac{d \sigma}{dS} =0 ,  \end{equation}
 \begin{equation}  \frac{d^{2} \mu_{y}}{d S^2}- -\frac{2}{\sigma} \frac{\mu_{y}}{dS} \frac{d \sigma}{dS} =0 ,  \end{equation}
 \begin{equation} \frac{d^{2} \sigma}{d S^2}- -\frac{1}{\sigma^2} ( \frac{d \sigma}{dS})^2  - \frac{1}{4 \sigma (r^2 -1)} [( \frac{d \mu_{x}}{dS})^2+ ( \frac{d \mu_{y}}{dS})^2 ] + \frac{r}{2 \sigma (r^2 - 1)}\frac{d \mu_{x}}{dS}\frac{d \mu_{y}}{dS}=0.  \end{equation}
  correlated systems $ r \neq 0$ one obtains
$$\mu_{x} =  -  \sqrt{ (\frac{2\\tilde{A}(r-1) }{\tilde{B}})} \tanh ( \frac{2AB}{2 r-1}S) , $$

$$\mu_{y} =  - \sqrt{(\frac{2A(r-1) }{B})}  tanh ( \frac{2AB}{2 r-1}S) , $$

$$\sigma  = - \sqrt{(\frac{-A }{B})}  sech ( \frac{2AB}{2 r-1} S) .  $$

\subsection{Chaotic Instability in Information Geometry}
It is well known that the Riemannian curvature of a manifold is closely connected with the  behavior of geodesics. If we take a special case   $\sigma_{x}= \sigma_{y} =\sigma$,
therefore , its corresponding Fisher-Rao metric becomes [14]
$$
g_{ij} =  \left (  \begin{array}{ccc}   \frac{1}{\mu^{2}_{1}}  & 0  &  0   \\    0 & \frac{1}{\mu^{2}_{2}} & 0  \\  0 & 0 &    \frac{1}{\sigma^{2}_{2}} \end{array} \right )
$$
to produce the line element
 \begin{equation}
  dS^{2}_{M^3D} = \frac{1}{\sigma^{2}}(d \mu^{2}_{x} + d \mu^{2}_{y}+ 4 d \sigma^{2}_{x} )  \end{equation}
and its  the non-vanishing affine connection becomes
$$ \Gamma^{1}_{13} = -\frac{1}{\sigma} , \Gamma^{2}_{23}= -\frac{1}{\sigma} = \Gamma^{1}_{32} ,$$   $$\Gamma^{3}_{11} =  \frac{-1}{{4 \sigma}} ,,  \Gamma^{3}_{22}=  \frac{-1}{{4 \sigma}} , \Gamma^{3}_{33} = - \frac{1}{ \sigma}. $$
The geodesic equation  describes a reversible dynamics whose solution is the trajectory between initial $\Theta_{i}$l and final macro-state $\Theta_{f}$  which can be expressed in the following way [15]
 \begin{equation} \frac{d^{2} \mu_{x}}{d S^2}- -\frac{2}{\sigma} \frac{\mu_{x}}{dS} \frac{d \sigma}{dS} =0 ,  \end{equation}
 \begin{equation}  \frac{d^{2} \mu_{y}}{dS ^2}- -\frac{2}{\sigma} \frac{\mu_{y}}{dS} \frac{d \sigma}{dS} =0 ,  \end{equation}
 \begin{equation} \frac{d^{2} \sigma}{d S^2}- -\frac{1}{\sigma^2} ( \frac{d \sigma}{dS})^2  - \frac{-1}{4 \sigma  } [( \frac{d \mu_{x}}{dS})^2 =0 .  \end{equation}
  If the Riemannian curvature is positive , then the nearby geodesics  oscillate about one another due to solution of geodesic deviation equations while when the curvature is negative,  the geodesics are rapidly diverge from each and the solution of geodesic deviation equations may give an indication about the behavior of this divergence. This provides a way to estimate the degree of chaotic behavior in the system , which means the estimate the  chaotic  issue.
i.e.
 \begin{equation} \frac{d^2 \Psi^{1}}{d S^2} + 2 \Gamma^{1}_{11}\frac{d \Theta^{1}}{(d S)^2} +\Gamma^{1}_{11}(\frac{d \Theta^{1}}{(d S)})^2  \Psi^{1} =0,  \end{equation}
 \begin{equation}  \frac{d^2 \Psi^{2}}{dS^2} +2[  \Gamma^{2}_{23} \frac{d \Theta^3}{dS}
\frac{d \Psi^2}{dS} + \Gamma^{2}_{32} \frac{d \Theta^2}{dS}\frac{d \Psi^3}{dS}] + \partial_{3} \Gamma^{2}_{23} ( \frac{d \Theta^3}{dS} ^2)(\Psi)^2 $$

$$~~~~ + \Gamma^{2}_{32}\Gamma^{3}_{33}( \frac{d \Theta^3}{d s})^2 \Psi^2 = \frac{1}{g_{22}} R_{2323} \frac{d \Theta^2}{dS} \frac{d \Theta^2}{dS}\Psi^3  +  \frac{1}{g_{22}} R_{2323} ( \frac{d \Theta^3}{dS})^2 \Psi^3 , \end{equation}
and
 \begin{equation}  \frac{d^2 \Psi^{3}}{dS^2} +2[  \Gamma^{3}_{23} \frac{d \Theta^3}{dS}
\frac{d \Psi^2}{dS} + \Gamma^{3}_{32} \frac{d \Theta^2}{dS}\frac{d \Psi^3}{dS}] + \partial_{3} \Gamma^{3}_{23} ( \frac{d \Theta^3}{dS} ^2)(\Psi)^2 $$

$$~~~~ + \Gamma^{3}_{32}\Gamma^{3}_{33}( \frac{d \Theta^3}{d s})^2 \Psi^2 = \frac{1}{g_{33}} R_{2323} \frac{d \Theta^2}{ds} \frac{d \Theta^2}{dt}\Psi^3  +  \frac{1}{g_{33}} R_{2323} ( \frac{d \Theta^3}{dt})^2 \Psi^3 . \end{equation}
After some manipulation,  the solution of geodesic equation and geodesic deviation equation may be expressed  as follows :
$$
\mu_{x} =  -  \sqrt{ (\frac{-2A }{B})} tanh ( - 2AB S) $$
$$
\mu_{y} =  - \sqrt{(\frac{- 2A }{B})}  tanh ( - {2AB}S) $$
$$  \sigma  = - \sqrt{(\frac{-A) }{B})} sech  ( {-2AB} S)  $$
and
$$ \Psi_{1} = ( a_{1} + a_{2}\rho)e^{ -r  \rho s} $$ ,   $$\Psi_{1} = ( a_{3} + a_{4}\rho)e^{ -  \rho s} - \frac{1}{2 \rho} a_{5}e^{ -  \rho s} + a_{6}$$, $$ \Psi_{3} = ( a_{3} +a_{4}\rho ) e^{- \rho s} $$
where, $a_{1}, a_{2}...\& a_{6}$  are integration constants and $\rho$ is a parameter defining the deviation vector such that $ \Psi^{i} = \frac{\partial x^{i}}{\partial \rho} $.  cf.(  Bazanski 1989)
allowing us to compute the chaotic behavior in the system using the scalar value of the deviation vector i.e.
$$
\Psi ^2 = \frac{1}{\mu_{1}^2} ( \Psi_{1})^2 + \frac{1}{\sigma_{2}^2} ( \Psi_{2})^2 + \frac{1}{\sigma_{2}^2} ( \Psi_{3})^2
$$
which becomes $$ \Psi = \bar{C} e^{\rho S }$$
where $\bar{C}$  is an arbitrary constant that encodes  information about the initial conditions and depends on the parameter ${\rho}$.
Thus, studying in depth some examples of systems whose data may be expressed as a statistical manifold having a negative curvature less than 1 may show how chaotic systems may be controlled.  This can be done through geometrization of the economic or financial system in order to maintain the risk in  the system within any limits assigned.
\section{Discussion and Concluding Remarks}
In this paper, we have suggested  a mechanism to express the differential equations of prey-predator model may be used to describe Capital-Income model into a space  expressing all of its factors as dimension in a geometric space by considering each of its elements as a dimension in a manifold.  Some authors have used an allometric space with a stochastic metric [3].  This could be applied as an introductory step to apply information geometry. This geometry has the advantage of expressing each individual data as micro-space each of which has its own macroscopic structure. In other words, we have described a mathematical technique for uniting microeconomics and macroeconomics. However, some problems are still existed in our current lives. The macrostructure produced by elements of microeconomics is not identical with that described in terms of space microstructure, described in terms of a space micro-structure is  not identical. Accordingly,   we may expect some current macroeconomic curves that are  controlling the effect of negative curvature from the background,  having  tendency of chaotic behavior appeared in the system.  From the perspective of economics, it may be considered a new tool for testing stable economy and its relation with  its corresponding  micro-economic  items. This geometrization may provide a tool to study the stability of the economy of countries experiencing rapid change due to the transitional situation of their economies. This descriptive study will be assigned for the forthcoming work.

\section*{Acknowledgements}
The author would like to thank Professors  K.Buchner, G. De Young, M.I.Wanas , M. Abdel Megied and his colleague Dr. E. Hassan for their remarks and comments to work in the field of multidisciplinary sciences.

\section{References}
{[1]} Cheryshov, S.I. , Voronin, A.V. and Ranzumovsky, S.A. (2009) ArXiv 0904.0756  \\
{[2]} Kahil, M.E. (2011)  WSEAS, Transaction on Mathematics, vol 10, 455 \\
{[3]} Dodson, Ci.J. (2009) 0903.2997 \\
{[4]} Marriot P., Salmon, M.  {\it{Applications of Differential Geometry to Econometrics}} ,   

 Cambridge University Press     pp. 190. \\
{[5]} Bazanski, S.L. (1989) J. Math Phys., {\bf{30}},1018. \\
{[6]} Kahil, M.E.  (2006)  J.Math. Phys  {\bf{47}}, 052501 \\
{[7] } Burda, Z.J. , Jurkiewicz, J. and  Nowak, M.A.(2003)  cond-mat/030109 \\
{[8]} Chouststva, Olga (2001) quantum-ph/0109122  \\
{[9]}  Zarikas, V., Christoploulos,A.G. and  Rendoumis, V. L. (2009) European  Journal of

 Economics, Finance, and Administrative Sciences, {\bf{16}}, 73. \\
{[10] }Dragulescu, A., and Yakovenko, V. M.  (2000) Eur. Phys. J.{\bf{B}},1723 \\
{[11]} Caticha, A. (2001) gr-qc/0109068  \\
{[12]} Kim, D.H., Ali, S.A. ,  Carfaro, C. and Mancini,S. (2011) ArXiv 1104.1250   \\
{[13]} Caticha, A. (2005)   gr-qc/0508108 \\
{[14] } Caticha, A. and  Cafaro, C. (2007)   AriXiv 0710.1071 \\
{[15]} Cafaro,A. and Ali, S.A.(2007) ArXiv n lin/ 0702027 

\end{document}